\documentclass[aps,pre,onecolumn,superscriptaddress,showpacs]{revtex4-1}
\usepackage{natbib}
\usepackage[english]{babel}
\usepackage[dvips]{graphics}
\usepackage{graphicx,epsfig}
\usepackage{amsmath}
\usepackage{color}
\usepackage{float}
\usepackage{multirow}
\usepackage[normalem]{ulem}
\usepackage{booktabs}
\usepackage{amsfonts} 
\usepackage{multirow}
\usepackage{subcaption}
\usepackage{braket}
\usepackage{caption}
\newcommand{\trans}{{\perp}}

\DeclareMathOperator{\Tr}{Tr}
\DeclareMathOperator{\arccosh}{arccosh}

\begin{document}
\title{
Bifurcations and phase-space structures in  KCN molecular system
}

\author{F. Revuelta}
\affiliation{Grupo de Sistemas Complejos, 
    Escuela T\'ecnica Superior de Ingenier\'ia Agron\'omica,
    Alimentaria y de Biosistemas,
    Universidad Polit\'ecnica de Madrid,
    Avenida\ Puerta de Hierro 2-4, 28040 Madrid, Spain.} 
\author{F. J. Arranz}
\affiliation{Grupo de Sistemas Complejos, 
    Escuela T\'ecnica Superior de Ingenier\'ia Agron\'omica,
    Alimentaria y de Biosistemas,
    Universidad Polit\'ecnica de Madrid,
    Avenida\ Puerta de Hierro 2-4, 28040 Madrid, Spain.} 
\author{R. M. Benito}
\affiliation{Grupo de Sistemas Complejos, 
    Escuela T\'ecnica Superior de Ingenier\'ia Agron\'omica,
    Alimentaria y de Biosistemas,
    Universidad Polit\'ecnica de Madrid,
    Avenida\ Puerta de Hierro 2-4, 28040 Madrid, Spain.} 
\author{F. Borondo}
\affiliation{Departamento de Qu\'imica, 
    Universidad Aut\'onoma de Madrid, 
    Cantoblanco, 28049 Madrid, Spain.}
%
%


\begin{abstract}
In this work,
we analyze the evolution of the phase-space structures of KCN molecular system
as a function of the vibrational energy using Lagrangian descriptors.
For low energies, 
the motion is mostly regular around the absolute minimum 
of the potential energy surface.
As the energy increases, 
the phase space combines regions with regular 
and chaotic motion,
a difference that is well captured by the Lagrangian descriptors.
We show that the dynamics is mostly governed by the invariant manifolds
of the stretch periodic orbits located at the top of one of the
energetic barriers of the system.
Furthermore, 
we show a perfect agreement between the bifurcation theory
and the differences observed in the phase-space structures
as the vibrational energy is modified.
The accuracy of our calculations is also assessed
by explicit comparison with the invariant manifolds computed 
using linear dynamics.
\end{abstract}


\maketitle

\begin{flushleft}
\end{flushleft}

\normalsize




\section{Introduction} \label{sec:intro}

Periodic orbits (POs) are 
trajectories that describe periodic cycles~\cite{Cvitanovic16}, i.\,e.,
they return to the same point after a time 
named as
the period of the PO.
Although typical POs themselves occupy a finite region of phase space,
they are surrounded by geometrical structures which can significantly
influence the system's global dynamics~\cite{Katok95, Celletti10}. 
On the one hand,
stable POs are resilient to disturbances and return to their original
path after suffering a small perturbation.
They are surrounded by invariant tori where quasiperiodic motion occurs.
On the other hand,
unstable POs exhibit, conversely, different behavior:
small perturbations increase exponentially over time.
Moveore,
they can be embedded within the chaotic sea
without presenting the 
usual
erratic motion shown by
generic chaotic trajectories. 
Finally,
marginally stable POs 
are on the boundary between stability and instability,
as small perturbations do not grow or shrink over time.

When a nonlinear system is manipulated by chaging some of its parameters
(such as energy, mass, shape,~\ldots),
the POs can undergo
dramatic changes.
Bifurcations~\cite{Hale91} can lead to the emergence of new POs,
thereby drastically altering the phase-space structure.
Changes in the stability of POs can profoundly impact system dynamics~\cite{Cvitanovic16};
for example, a stable PO may become marginally stable and eventually unstable,
or an unstable PO can become, first,
marginally stable, and then stable.
Extensive literature, spanning from nonlinear oscillators~\cite{Lakshmanan96}
to fluid dynamics~\cite{Kambe04}
and celestial mechanics~\cite{Celletti10},
illustrates these phenomena.
At the same time,
when bifurcations occur,
the invariant geometrical objects responsible for the
dynamics can be distorted, disappear, or new ones may emerge. 

Several tools have been developed over time
in order to analyze 
phase-space structures.
In the case of dynamical systems with 2 degrees of freedom (dofs),
the Poincaré surface of section (PSS) remains a cornerstone~\cite{LL10}.
This tool allows the unambiguous identification of invariant tori
and efficiently determines whether motion is regular or chaotic.
The Lyapunov exponent is another remarkable chaos indicator 
also applicable 
in systems with more than 2 dofs~\cite{Cvitanovic16}.
However,
its calculation is computationally demanding.
Therefore, 
alternatives have been developed,
such as the
fast Lyapunov indicator~\cite{Froeschle97, Froeschle00},
the small-alligment index~\cite{Skokos01, Benitez15}
and 
the mean exponential growth factor of nearby orbits~\cite{Cincotta00, Cincotta03},
among others.
Lagrangian descriptors (LDs)~\cite{JM10, Lopesino15, Mancho13, Mendoza10}
also offer an attractive
approach to unveil the underlying phase-space structures.
They have been successfully applied in various fields
such as
oceanic~\cite{Mendoza10}
and
cardiovascular flows~\cite{Darwish21},
chemical~\cite{Craven15, Pandey22}
and
molecular systems~\cite{Forlevesi23, Revuelta19, Revuelta21, Revuelta23},
and
chaotic maps~\cite{Carlo20, Lopesino15, SB24, Zimper23}
and billiards~\cite{Carlo22}.
These major tools, along with others,
enable the study of
the broad field of nonlinear dynamics,
forming the core of chaos theory.
Furthermore,
they can be applied to study molecular vibrations
as vibrational motion takes place
when atomic particles sojourn 
the potential energy surface (PES)
exerted by the electronic structure~\cite{Arnaut21, Atkins06, Wales03}.
This is especially interesting
when the vibration energy is well above
the ground-state energy
since then
the non-negligible anharmonic terms in the PES
make
normal mode analysis unsuitable~\cite{Ma05}.


The standard description of molecular systems typically lies on the 
Born-Oppenheimer or adiabatic approximation~\cite{Atkins06, Wales03}.
Within this approach,
the motion of the atomic nuclei 
is separated from that of the electrons,
due to their significant mass difference,
which spans at least three orders of magnitude.
Consequently,
the PES
associated with the electrons
is computed for each position of the atomic nuclei.
The  computation of the PES 
often
requires
the application of advanced 
\emph{ab-initio} quantum chemistry tools~\cite{Marx09},
something usually 
very time-consuming and
computationally demanding.
The atomic nuclei 
undergo
vibrational motion on the PES
created by the electrons.
For small energies, 
atomic nuclei simply vibrate around the equilibrium positions
found at the PES minima.
However, 
when the energy is sufficiently high,
nuclei can explore larger regions of the PES,
rendering other more complex phenomena such as isomerization.
This structural change in the molecule
can have significant
 implications in its chemical properties and stability~\cite{Atkins06, Wales03}.
The typically strong anharmonicities
in the PES can make this process extraordinary 
complicated.
Nevertheless,
concepts and tools  developed within the fields
of nonlinear mechanics and dynamical systems theory
can greatly facilitate the study~\cite{Uzer02}.
For example, 
they can be used to establish whether nuclei
present regular or irregular motion.
Moreover,
they can be applied to identify the underlying geometrical structures that
determine nuclear motion,
as well as to identify the reaction mechanism that
is responsible, among others,
for molecular isomerization~\cite{Uzer02}.
Following this mathematical perspective,
in configurations close to the PES minima,
nuclei simply vibrate because of the existence of
invariant tori which force the motion to be regular
within a certain region of phase space.
As a consequence,
the nuclei can show periodic or quasiperiodic motion,
depending on where they are located on a regular PO
or not.
As the energy increases,
according to the Poincaré-Birkhoff theorem~\cite{Birkhoff13},
the occurrence of bifurcations leads
to
the appearance of new POs,
while invariant tori break down,
as prescribed by the
Kolmogorov-Arnold-Moser theorem~\cite{Arnold06, LL10}.
Consequently,
the route for chaotic motion is opened,
making the dynamics much more intricate.
Still,
just like in any other generic nonlinear system,
stable and unstable POs can be found,
corresponding to situations where the nuclei 
exhibit
periodic motion.
Moreover,
the motion of the nuclei when located in the vicinity of 
these objects can be also well understood.
On the one hand,
when located close to a stable PO,
the
nuclei describe quasiperiodic vibrational motion
due to the existence of invariant tori.
On the other hand,
when found in the vicinity of an unstable PO,
the motion of nuclei is more involved but it can be
accurately described within a linear approximation.
Nuclei exponentially separate for the reference PO
in the directions of the unstable manifolds.
At
larger distances,
the manifolds fold and
the linear approximation breaks down,
but the importance of the invariant manifolds
is still substantial in the system dynamics.

In this work,
we 
investigate
the phase-space structures 
responsible
for the intricate 
nonlinear dynamics of KCN molecular system
using LDs.
For very low energies,
we show the existence of a single regular PO
associated with the vibrational motion of K atomic nuclei
around the
absolute minimum of the PES.
As the energy increases,
the primary PO experiences a series of (infinite) bifurcations.
The most dramatic one occurs at a relative low energy,
where 
the PO losses stability,
triggering the emergence of another stable PO.
This newly formed stable PO
remains stable within a reasonably
small range of energy until another bifurcation arises.
For
higher
energies,
the global dynamics
are primarily
governed by the invariant manifolds
that emerge from unstable POs
located at the top of one of the energetic barriers.
This phenomenon is
well captured by the LDs.
This secondary PO also suffers a series of bifurcations,
which we similarly analyzed.
We discuss the corresponding changes in the phase-space structures,
with special attention given to the unexpected stabilization
process of the PO as the energy is modified.
At
a certain bifurcation energy,
the secondary PO transforms into a stable orbit,
resulting in regular motion of K atomic atomic nuclei
within a small region of phase space.
This stabilization process is well captured by the LDs,
and is in agreement with dynamical systems theory.
We illustrate that the bifurcation occurs
when the stable and unstable manifolds of the 
reference POs are distorted and degenerate
at the bifurcation energy.
As the energy increases,
the size of the stability region, 
which is associated with invariant tori,
first increases, and then reduces until it dissappears,
causing the PO to return to an unstable state.
We briefly discuss some of the bifurcations that occur in between,
demonstrating excellent agreement between the LDs
computations and dynamical systems theory.
To conclude,
in order to strengthen the accuracy of LDs,
we explicitly compare the LDs results with
numerically-computed invariant manifolds.

The 
structure
of the paper is as follows.
Following this introduction,
we provide a brief overview of the KCN molecular system
under study in Sec.~\ref{sec:system}.
In Sec.~\ref{sec:methods},
we introduce
the methods and tools used to analyze the
system.
Section~\ref{sec:results} presents the
results obtained and the corresponding discussion.
Finally,
we conclude the article with a summary
in Sec.~\ref{sec:summary}.


\section{System} \label{sec:system}

\begin{figure}[t!]
\centering
\includegraphics[width=0.95\columnwidth]{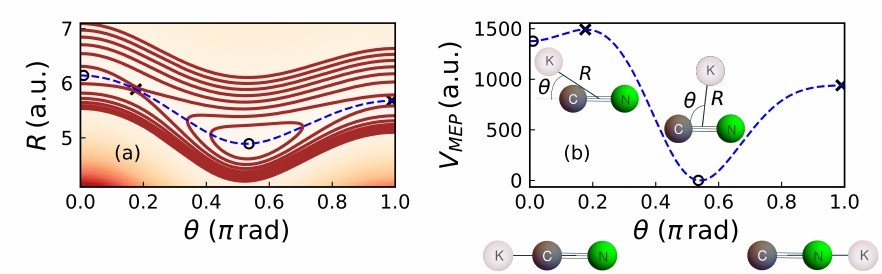}
\caption{KCN molecular system.
(a) Fundamental domain of the \emph{ab-initio} potential energy surface 
as a function of the Jacobi coordinates~$R$ and~$\theta$.
The continuous brown lines show the equipotential lines for
$V(R, \theta) = 500, 1000, \ldots, 3500, 4000$\,cm$^{-1}$.
The potential minima (black circles) and the saddle points (black crosses)
are joined by the superimpossed minimum energy path (dashed blue line).
(b) Profile of the potential energy surface along the minimum energy path
(dashed blue line) with the corresponding critical points (black symbols).
Insets: Sketches of the configuration of
KCN molecule at the left saddle point and at the potential minimum
with the Jacobi coordinates.
Bottom: Sketches of the colinear configurations of the molule for~$\theta = 0$\,rad (left),
and~$\theta = \pi$\,rad (right).
}
\label{fig:KCN}
\end{figure}

In this work,
we study the vibrational motion of
potassium cyanide molecular system, KCN.
In the rotationless setting, 
the configuration of this triatomic molecule can be 
characterized using the Jacobi coordinates
shown in the insets of Fig.~\ref{fig:KCN}(b).
Here,
$R$ is the 
radial distance 
between the potassium atom and the center of mass
of the CN group,
and~$\theta$ is the angle that the previous radius forms
with the 
longitudinal axis of CN.
Due to the triple bond that bounds carbon and nitrogen atoms,
this vibrational mode 
can be effectively decoupled
from the rest of the molecule.
Thus, in practice,
the C-N distance can be set equal to the equilibrium value~$r_{\rm eq} = 2.224$\,a.u.
Then,
the system can be accuratelly described by using the following
2-dofs
Hamiltonian
\begin{equation}
   \mathcal{H}(P_R, P_\theta, R, \theta) = \frac{P_R^2}{2 \mu_1} +
 \frac{P_\theta^2}{2} \left( \frac{1}{\mu_1 R^2} +  \frac{1}{\mu_2 r_{\rm eq}^2}  \right)
+V(R, \theta),
\end{equation}
where~$P_R$ and~$P_\theta$ are the conjugate momenta to~$R$ and~$\theta$
coordinates,
$\mu_1 = m_K (m_C + m_N) / (m_K +m_C + m_N) $
and
$\mu_2 = m_C m_N/ (m_C + m_N) $
are the reduced masses for K-CN and CN groups, respectively,
and~$V(R, \theta)$ is the Bohr-Oppenheimer PES
that determines the surrounds of the atomic nuclei 
due to the electronic structure.

Figure~\ref{fig:KCN}(a) shows as a contour plot
the fundamental domain ($0 \le \theta \le \pi$\, rad)
of the \emph{ab-initio} PES
taken
from Ref.~\cite{Parraga13}.
As can be seen,
the PES presents two minima (black circles) and two saddle points
(black crosses),
which are joined by the minimum energy path (MEP) 
$R_{\rm MEP}(\theta)$, plotted
superimpossed as a dashed blue line.
These critical points are more clearly visible in 
Fig.~\ref{fig:KCN}(b),
where the profile of the PES along the MEP is represented.
For~$\theta=0$ and~$\pi$\, rad,
respectively,
the molecule forms the colinear configurations 
K-CN (left) and CN-K (right)
shown at the bottom of Fig.~\ref{fig:KCN}(b).
The configurations for the intermediate saddle and minimum points
are also sketched in the insets of Fig.~\ref{fig:KCN}(b).

\section{Methods} \label{sec:methods}
In this section, we report on the methods used to conduct our study.
First, we summarize in Sec.~\ref{sec:LD} the tool used to identify
the phase-space structures that are responsible for the dynamics of
the molecule under study: the LDs.
Second, we briefly discuss the linearized dynamics around
periodic orbits in Sec.~\ref{sec:linear_motion},
and discuss the changes in stability as a function of the energy,
including the occurrence of bifurcations.

\subsection{Lagrangian descriptors}\label{sec:LD}
The tool selected to analyze the phase-structures that are responsible
for the highly nonlinear dynamics of KCN are LDs.
For a system with $N$ dofs,
LDs are defined as~\cite{Lopesino15}
\begin{equation} \label{eq:LD}
  M_{\pm} = \pm \sum_{i=1}^{2N} \int_0^{\pm \tau} \vert \dot z_i(t) \vert^p \textnormal{d}t,
\end{equation}
with~$0<p \le 1$, and~$\mathbf{z}(t)$ a symplectic coordinate,
that solves the equations of motion
\begin{equation}
   \dot z_i = \mathbf{J} \frac{\partial \mathcal{H}}{\partial z_i}, 
\qquad
\textnormal{with}
\quad
\mathbf{J} = \left(
\begin{array}{cc}
   \mathbf{0}_{N \times N} & \mathbf{I}_{N \times N} \\
  -\mathbf{I}_{N \times N} & \mathbf{0}_{N \times N}
\end{array}
\right) . \label{eq:EoM}
\end{equation}
As will be shown below 
in section~\ref{sec:results},
$M_-$, i.e.,
the  integration backwards in time of Eq.~\eqref{eq:LD},
unveils the unstable manifolds of a dynamical system,
whilst
$M_+$, i.e.,
the  integration forward in time,
unravels the stable manifolds.
Thus,
the sum $M_- + M_+$ can be used to identify 
the homoclinic and heteroclinic tangle
as it accounts for the stable and unstable manifolds at the same time.

For the case study, $N=2$
and~$\mathbf{z}(t) = \left( R(t), \theta(t), P_R(t), P_\theta(t) \right)$.
We have already shown elsewhere~\cite{Revuelta23} that a value of~$p=0.4$
and an integration time of~$\tau = 2 \times 10^4$\,a.u.
seem to be adequate for triatomic molecular systems~\cite{Revuelta19, Revuelta21}
like the one under study.

\subsection{The linearized motion}\label{sec:linear_motion}
The motion in the close vicinity of a PO can be 
described within a linear approximation~\cite{Cvitanovic16}.
Then, for
given a reference PO~$\mathbf{z}_{\rm PO}(t)$
with the initial condition~$\mathbf{z}_{\rm PO}(0)$,
the time evolution of an
initial condition~$\mathbf{z}(0) = \mathbf{z}_{\rm PO}(0) + \delta \mathbf{z}(0)$,
with~$\delta \mathbf{z}(0)$ sufficiently small,
can be characterized by the time evolution of
the vector~$\delta \mathbf{z}(t)$ that accounts for the distance to 
the reference PO, 
which is given by
\begin{equation}
    \delta \mathbf{z}(t) = \mathcal{M}(t) \delta \mathbf{z}(0),
\end{equation}
where~$\mathcal{M}(t)$ is a Jacobian matrix that accounts for the
linearized flow along the PO and satisfies that~$\mathcal{\dot M}(t) = \mathcal{L} \mathcal{M}(t)$,
with~$\mathcal{L}  = \mathbf{J} \frac{\partial^2 \mathcal{H}}{\partial z^2}$,
being~$\mathbf{J}$ given by Eq.~\eqref{eq:EoM},
and~$\mathcal{M}(0)=\mathbf{I}$.
Furthermore,
due to Liouiville theorem~\cite{LL10},
the determinant of the monodromy matrix always equals one.

After one period of time~$\mathbf{z}_{\rm PO}(T) = \mathbf{z}_{\rm PO}(0)$,
the Jacobian matrix~$\mathcal{M}(T)$
 is known as the {\it monodromy matrix} of the PO.
The dimension of this matrix is equal to the number of dofs.
Nonetheless,
all the nontrivial information
is solely portrayed by 
the transversal components to the PO.
For the case study, 
these components define a plane.
The corresponding \emph{transveral monodromy matrix}~$\mathcal{M}_1^\trans(T)$
has, as a consequence, dimension~2.

\subsubsection{Stability of periodic orbits
and bifurcations}\label{sec:bifur}

Since 
the determinant of~$\mathcal{M}_1^\trans(T)$
equals one,
the corresponding eigenvalues can be written as
\begin{equation} \label{eq:mu}
  \mu_\pm = \frac{\Tr\left( \mathcal{M}_1^\trans \right)}{2} \pm i \sqrt{1- \left( \frac{\Tr\left( \mathcal{M}_1^\trans \right)}{2} \right)^2},
\end{equation}
so that~$\mu_+ \mu_- = 1$, and~$\mu_+ + \mu_- = \Tr\left( \mathcal{M}_1^\trans \right) $
is a real number.

For stable POs,
the two eigenvalues are complex conjugates on the unit circle~$\mu_\pm = e^{\pm i \sigma}$, with~$\sigma \in \mathbb{R}$,
and then~$\Tr\left( \mathcal{M}_1^\trans \right) = 2 \cos \sigma$.
In this case,
$\vert \Tr\left( \mathcal{M}_1^\trans \right) \vert < 2$.
For unstable POs,
the two eigenvalues are real reciprocals~$\mu_+ = \mu_-^{-1}$,
and then~$\vert \mu_\pm \vert = e^{\pm \sigma}$.
We have,
as a consequence,
that
$\vert \Tr\left( \mathcal{M}_1^\trans \right) \vert = 2 \cosh \sigma$,
and~$\vert \Tr\left( \mathcal{M}_1^\trans \right) \vert > 2$.
When~$\mu_+>1$,
both eigenvalues and the trace are positive;
In this situation,
the unstable PO is  \emph{hyperbolic}.
Contrarily,
when~$\mu_+<-1$,
both eigenvalues and the trace are negative,
and the PO is called \emph{hyperbolic with reflection}.
In any case,
the inverse of the Floquet exponents
defined as
$\lambda_\pm =  \frac{\pm \sigma}{T}$
determines the characteristic time scale
for the linearized motion.
Finally,
when the two eigenvalues of~$\mathcal{M}_1^\trans$
are equal~$\mu_+ = \mu_- =  \pm1$,
and
$\vert \Tr\left( \mathcal{M}_1^\trans \right) \vert= 2$.
In this limiting case, a \emph{marginally stable PO} is found~\cite{Borondo24}.

For POs of $n$-period,
with~$n=1, 2, \ldots$,
the previous reasoning remains valid by substitution of~$\mathcal{M}_1^\trans$
by ~$\mathcal{M}_n^\trans = \left( \mathcal{M}_1^\trans \right)^n$.
Then, 
$\Tr\left( \mathcal{M}_n^\trans \right) = 2 \cos (n \sigma) =
2 \cos \left[ \arccos \left(  \Tr\left( \mathcal{M}_1^\trans \right) / 2 \right) \right]$,
or
$\Tr\left( \mathcal{M}_n^\trans \right) = 2 \cosh (n \sigma) =$
\linebreak
$ 2 \cosh \left[ \arccosh \left(  \Tr\left( \mathcal{M}_1^\trans \right) / 2  \right) \right]$,
depending on whether the PO is
stable or unstable, respectively.
When a bifurcation takes places,
a new $n$-period PO emerges or dissapears,
something that requires that
$\vert \Tr\left( \mathcal{M}_n^\trans \right) \vert = 2 $,
a condition that is fulfilled when
\begin{equation} \label{eq:bifur}
\Tr\left( \mathcal{M}_1^\trans \right) = 2 \cos \left( \frac{2 \pi m}{n} \right),
\end{equation}
with~$m$ an integer such that the cosine in Eq.~\eqref{eq:bifur} is~$\pi$-modulus.
Let us conclude by remarking that
unstable POs do not bifurcate,
i.e., 
they do dot render new POs,
but they can change their stability from
unstable to stable 
when the trace of their corresponding transversal
matrix equals~2.

In summary,
the eigenvalues of~$\mathcal{M}_1^\trans(T)$
given by Eq.~\eqref{eq:mu}
encode all the information
necessary to stablish whether a PO is stable,
unstable
or marginally stable,
and bifurcations occur when
 Eq.~\eqref{eq:bifur} is fulfilled.

\section{Results}
\label{sec:results}

This section describes the main findings of our study.
For this purpose,
we show
plots with the LDs computed for pairs of initial conditions
$(\vartheta, P_\vartheta)$
at the
PSS~\cite{LL10}
defined along the MEP
(MEP-PSS)~\cite{Parraga18},
where~$R=R_{\rm MEP}(\theta)$,
as
\begin{subequations} \label{eq:PSS}
	\begin{align}
   \vartheta &= \theta , \\
   P_\vartheta &= P_\theta   + P_R \frac{\textnormal{d}R_{\rm MEP}(\theta)}{\textnormal{d}\theta},
	\end{align}
\end{subequations}
for~$P_R<0$.
The MEP-PSS
is the most likely,
then useful, 
PSS
to be crossed
by a generic trajectory
as the MEP is the lowest energetic path that connects the system minima.
Thus,
in what follows we will compute the LDs
as defined in Eq.~\eqref{eq:LD}
for uniform sets of initial conditions located
on the MEP-PSS
with different vibrational energies.
Next,
we present in Secs.~\ref{sec:lowE} 
and~\ref{sec:highE}
the results for low and 
high vibrational energies, respectively.

\subsection{Phase-space structures for low energies}
\label{sec:lowE}

\begin{figure}[t!]
\centering
\includegraphics[width=0.95\columnwidth]{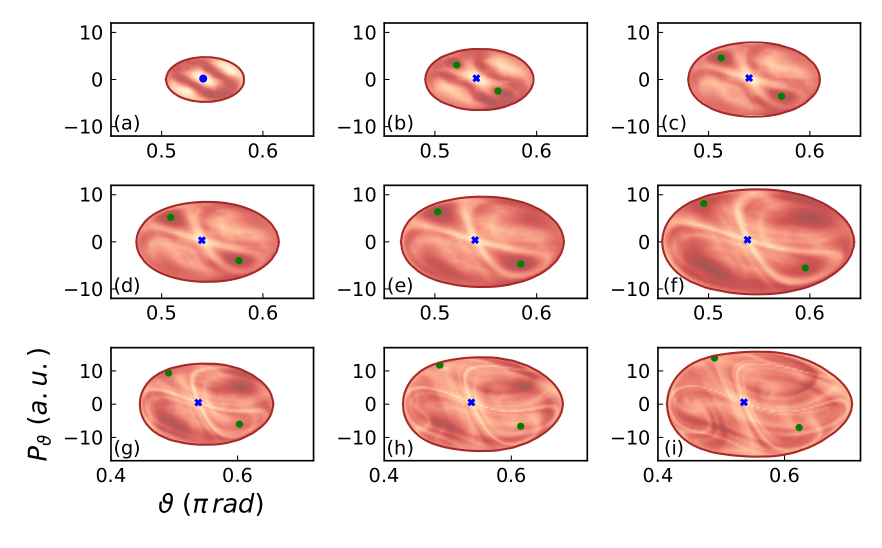}
\caption{Sum~$M_- + M_+$ of the Lagrangian descriptors~\eqref{eq:LD}
for~$p=0.4$ and~$\tau = 2 \times 10^4$\,a.u.,
and vibrational energies equal to
(a) $45$\,cm$^{-1}$,
(b) $85$\,cm$^{-1}$,
(c) $125$\,cm$^{-1}$,
(d) $145$\,cm$^{-1}$,
(e) $185$\,cm$^{-1}$,
(f) $250$\,cm$^{-1}$,
(g) $300$\,cm$^{-1}$,
(h) $400$\,cm$^{-1}$,
and
(i)~$500$\,cm$^{-1}$.
The blue  symbols mark the positions of the elliptic (circle) and
hyperbolic (cross) fixed points associated
with the 1:1 resonance,
and 
the green circles
those of the elliptic fixed points related to the 1:2 resonance.
%
}
\label{fig:LDs_smallE_t2d4}
\end{figure}

Figure~\ref{fig:LDs_smallE_t2d4} shows the LDs
defined in Eq.~\eqref{eq:LD}
for a uniform set of initial conditions
at the MEP-PSS.
As in Ref.~\cite{Revuelta23},
the value of the integration time has been set
equal to~$\tau = 2  \times 10^4$\,a.u.
(see discussion below on its appropriateness).
As expected,
the energetically accesible region of phase space
increases with the energy,
and so does the accessible area of the MEP-PSS.
For the lowest-lying energy considered,
$E=45$\,cm$^{-1}$, 
the LDs shown in Fig.~\ref{fig:LDs_smallE_t2d4}(a)
are slow varying functions,
which shows the existence of
regular motion.
Notice the \emph{donut}-like structure
around the central point
that is recognizable
in the plot.
This structure corresponds to a series of invariant
tori that separate the motion within the inner region from the motion 
within the outer part.
Notice the blue circle 
in the center of the plot
$(\vartheta, P_\vartheta) = (0.541, 0.196)$\,($\pi$\,rad,\,a.u.),
which marks the position of 
the elliptic fixed point associated with a primary stretch K-CN PO.
This trajectory is stable and has~$\theta(t) \approx \vartheta_{\rm PO} = 0.541\pi$\,rad.
Furthermore,
its radial, and angular frequencies,~$\omega_R$ and~$\omega_\theta$,
are equal,
indicating
 a $\omega_R : \omega_\theta = $ 1:1 resonance.

\begin{figure}[t!]
\centering
\includegraphics[width=0.95\columnwidth]{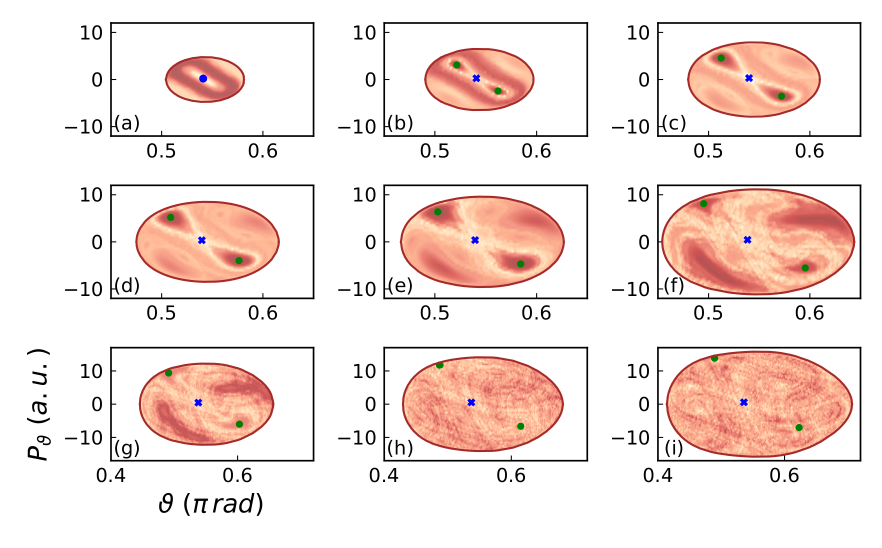}
\caption{
Same as Fig.~\ref{fig:LDs_smallE_t1d5}
for~$\tau =  10^5$\,a.u.
}
\label{fig:LDs_smallE_t1d5}
\end{figure}

For $E=45.9$\,cm$^{-1}$, 
the previous stable PO
bifurcates,
resulting
a central unstable PO and a neighboring stable PO.
As the energy increases,
the stable PO separates from the central region.
Figure~\ref{fig:LDs_smallE_t2d4}(b)
shows the LDs for $E=85$\,cm$^{-1}$, 
an energy where the two POs have already separared
substantially from each other.
Here, the blue cross located at
$(0.540, 0.264)$\,($\pi$\,rad,\,a.u.),
marks the position of hyperbolic point of the 
primary PO, which has turned unstable.
The green circles located at
$(0.521, 3.096)$\,($\pi$\,rad,\,a.u.),
and
$(0.562, -2.436)$\,($\pi$\,rad,\,a.u.),
correspond to the positions of the elliptic fixed points of the 
newly emerged stable PO.
This secondary PO corresponds
to a 1:2 resonance ($\omega_R = 2 \omega_\theta$).
Each of the correponding fixed points
is surrounded by 
a stability region, where the LDs are
slow-varying functions,
appearing
as almost uniform colored sectors
on the LD plots.
This stable chain of islands
is more clearly visible,
when the integration time is increased,
as concluded by comparison of 
Figs.~\ref{fig:LDs_smallE_t2d4}
and~\ref{fig:LDs_smallE_t1d5},
where a computational time 5 times larger has been considered
($\tau = 10^5$\,a.u.).
This time is more adequate at these low energies because 
the integration time of Eq.~\eqref{eq:LD}
must be larger than the
inverse of the Floquet exponent~$\lambda_\pm^{-1}$
of the  trajectory of interest~\cite{Revuelta21},
and then~$\tau =  2 \times 10^4$\,a.u. seems not to be sufficient.
In the center of 
Fig.~\ref{fig:LDs_smallE_t1d5}(b)
the hyperbolic point related to the unstable 1:1 PO is located,
whose invariant manifolds surround the previous chain of islands.
Notice, however, that the hyperbolic character of this PO is
only noticeable in its close vinicity,
as the distorted tori that surround the 1:1 and 1:2 POs
are still present,
effectively acting as dynamical barriers.
Notice, also, the bright points around the hyperbolic region
of Fig.~\ref{fig:LDs_smallE_t1d5}(b),
which correspond to a new set of elliptic points
that emergence because of a the occurrence of 
other bifurcations, 
as prescribed by the Poincaré-Birkhoff theorem~\cite{LL10}.

For $E=125$\,cm$^{-1}$, 
the results of Figs.~\ref{fig:LDs_smallE_t2d4}(c) and~\ref{fig:LDs_smallE_t1d5}(c)
show that 
the elliptic green fixed points 
move further appart from the central hyperbolic
blue cross.
Furthermore,
the size of the 
stability islands increase.
Conversely,
the size of the outer invariant tori decreases,
as prescribed by the
Kolmogorov-Arnold-Moser theorem~\cite{Arnold06, LL10}.
%
For $E=145$\,cm$^{-1}$, 
from visual inspection of 
Figs.~\ref{fig:LDs_smallE_t2d4}(d) and~\ref{fig:LDs_smallE_t1d5}(d),
one concludes that
the size of the stable islands
has noticeably increased.
%
For $E=185$\,cm$^{-1}$, 
the LD plots shown in Figs.~\ref{fig:LDs_smallE_t2d4}(e) and~\ref{fig:LDs_smallE_t1d5}(e)
present many more details.
In this case, 
the LDs change more significantly because of the presence
of invariant manifolds filling the 
region of phase space where the motion is chaotic.

For a vibrational energy of $E=250$\,cm$^{-1}$,
as considered in Figs.~\ref{fig:LDs_smallE_t2d4}(f) and~\ref{fig:LDs_smallE_t1d5}(f),
new structures
can be recognized.
The size of the region presenting chaotic motion
has further increased, 
while the size of the regions of regular motion,
characterized by a large value in the LDs
and a dark color in the plots, 
reduces.
This behavior is also observed for
$E=300$\,cm$^{-1}$
in Figs.~\ref{fig:LDs_smallE_t2d4}(g) and~\ref{fig:LDs_smallE_t1d5}(g),
$E=400$\,cm$^{-1}$
in Figs.~\ref{fig:LDs_smallE_t2d4}(h) and~\ref{fig:LDs_smallE_t1d5}(h),
and
$E=500$\,cm$^{-1}$
in Figs.~\ref{fig:LDs_smallE_t2d4}(i) and~\ref{fig:LDs_smallE_t1d5}(i), 
respectively.
In all these cases,
the dynamics in most of the accessible phase space
is dominated by the invariant manifolds
that emerge from the central 1:1 resonance.
Notice that for all these higher energies, 
as well as other larger ones,
the integration time
of~$\tau = 2 \times 10^4$\,a.u. considered in Fig.~\ref{fig:LDs_smallE_t2d4}
seems to be more adequate than that
of~$\tau = 10^5$\,a.u. used in Fig.~\ref{fig:LDs_smallE_t1d5}
to identify the invariant manifolds that emerge from the
central stretch PO.
The reason lies on the stability of this PO,
which has a larger characteristic stability exponent,
so its inverse is smaller.

\subsubsection{Bifurcation scheme for low energies}
\label{sec:bifur_lowE}

\begin{figure}[t!]
\centering
\includegraphics[width=0.95\columnwidth]{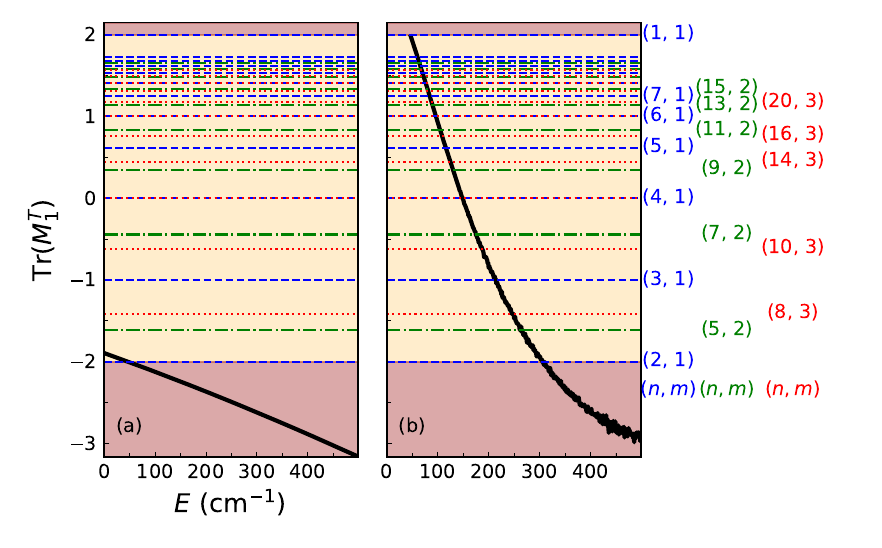}
\caption{Trace of the monodromy matrix as a function of the vibrational energy
for 
(a) the central 1:1 periodic orbit,
and
(b) the outmost 1:2 periodic orbit.
The orange central and the redish regions
correspond to stable and unstable behavior, respectively.
The horizontal lines mark some of the positions where the bifurcation condition
given by Eq.~\eqref{eq:bifur} is satisfied for
$n=1$ (dashed blue),
$n=2$ (dotted-dashed green),
~$n=3$ (dotted red).
The first~$(n, m)$ values 
appear on the right-hand side of the plot.
The corresponding bifurcation energies are 
 listed in Table~\ref{table:bifur}.
}
\label{fig:tr_centalPOs}
\end{figure}

To provide a more comprehensive qualitative analysis of
phase-space structural changes at low energies, 
we show in Fig.~\ref{fig:tr_centalPOs}
the trace value of the 
transversal monodromy matrix~$\mathcal{M}_1^\trans(T)$
with respect to the vibrational energy,
for both
(a) the central 1:1 resonance,
and
(b) the outermost 1:2 resonance.
Here,
the horizontal lines highlight some of the values 
where the trace fulfills Eq.~\eqref{eq:bifur}
for~$n=1$ (dashed blue),~$n=2$ (dotted-dashed green),
and~$n=3$ (dotted red).
Some of the $(n, m)$ pairs can be found on the
right-hand side of the figure.
These pairs of numbers approximately correspond to the
quantization numbers of the
most regular molecular eigenstates~\cite{Parraga13}.
The combinations that yield the same trace value
as those involving lower integers,
e.\,g. $(n, m) = (4, 2)$ (equivalent to~$(n, m) = (2, 1)$)
and~$(n, m) = (9, 3)$ (equivalent to~$(n, m) = (3, 1)$),
have been omitted.

Initially,
we observe 
that the 1:1 resonance is stable
for vibrational energies below
the bifurcation threshold~$E_{\rm bifur, 1}^{\rm min} = 45.9$\,cm$^{-1}$,
where a bifurcation~$(n, m) = (1, 1)$ takes place.
Conversely, energies surpassing~$E_{\rm bifur, 1}^{\rm min}$,
render the 1:1 resonance unstable,
as the trace value  
monotonically decreases with the energy
(at least within the range considered).
 We identify the corresponding invariant manifolds
though LDs,
visually discerned in
Figs.~\ref{fig:LDs_smallE_t2d4}(e)-\ref{fig:LDs_smallE_t2d4}(i)
and~\ref{fig:LDs_smallE_t1d5}(b)-\ref{fig:LDs_smallE_t1d5}(d).

Second, when the vibrational energy equals
$E_{\rm bifur, 1}^{\rm min}$,
the previous 1:1 resonance becomes marginally stable.
Then,
an additional 1:2 resonance 
degenerated with the 1:1 resonance emerges.
The trace value of the corresponding
transversal monodromy matrix
also decreases on average with some small oscillations.
The 1:2 resonance remains stable 
(orange central region)
for energies below
$E_{\rm bifur, 2}^{\rm min} = 308.4$\,cm$^{-1}$
and unstable for energies above
$E_{\rm bifur, 2}^{\rm min}$
(redish region),
where a  bifurcation~$(n, m) = (2, 1)$ occurs.
This finding is consistent with the
presence of the stable islands 
in Figs.~\ref{fig:LDs_smallE_t2d4}(b)-\ref{fig:LDs_smallE_t2d4}(g)
and~\ref{fig:LDs_smallE_t1d5}(b)-\ref{fig:LDs_smallE_t1d5}(g),
and with their absence
in Figs.~\ref{fig:LDs_smallE_t2d4}(h)-\ref{fig:LDs_smallE_t2d4}(i)
and~\ref{fig:LDs_smallE_t1d5}(h)-\ref{fig:LDs_smallE_t1d5}(i).
%
The 1:2 resonance also experiences infinite series of
bifucations within the range
$(E_{\rm bifur, 1}^{\rm min}, E_{\rm bifur, 2}^{\rm min})$.
Some examples of the corresponding bifurcating energies
are listed in Table~\ref{table:bifur}.

\begin{table}[t!]
\centering
\caption{Energies in cm$^{-1}$,
where the bifurcation condition
\eqref{eq:bifur}
for~$n$-period periodic orbits with~$n=1, 2, \ldots, 10$
(left)
for the  resonance 1:2 located close to the absolute minimum
of the potential,
and
(right)
for the resonance 1:1 that is localized close to the saddle point
located at the top of the energetic barrier~$\theta = \pi$\,rad. 
All accessible values of~$m$ are considered.
The positions marked with a hyphon ($-$)
provide the same results as other combinations of $(n, m)$.
}
\label{table:bifur}
\setlength{\tabcolsep}{22pt}
\footnotesize{
\begin{tabular}{ccccccc}
\hline\hline\hline
& \multicolumn{3}{c}{$m$ (Resonance 1:2 close to the minimum)} 
& \multicolumn{3}{c}{$m$ (Resonance 1:1 close to the right saddle point)} \\
$n$ & 1 & 2 & 3  & 1 & 2 & 3  \\
\hline
  1 &   45.9 & $-$ & $-$ & 1343.2 & $-$ & $-$ \\
  2 & 308.4  & $-$ & $-$ & 4220.6 & $-$ & $-$ \\
  3 & 211.8  & $-$ & $-$ & 3453.5 & $-$ & $-$ \\
  4 & 148.1  & $-$ &  $-$ & 2723.7 & $-$ & - \\
  5 & 115.7  & 262.7 & $-$ & 2287.5 & 3922.5 & $-$ \\
  6 & 96.5  & $-$ & $-$  & 2022.6 & $-$ & $-$ \\
  7 & 84.3 & 174.6 & 286.1 & 1853.0 & 3044.5 & 4065.2   \\
  8 & 76.0  & $-$ & 246.2 & 1738.9 & $-$ & 3766.1 \\
  9 & 70.1 &  129.9 & $-$ & 1658.8 & 2477.3 & $-$ \\
10 & 65.8  & $-$ & 187.0 & 1600.5 & $-$ & 3170.8\\
%
\hline\hline\hline
\end{tabular}}
\end{table}

\subsection{Phase-space structures for high energies}
\label{sec:highE}

\begin{figure}[t!]
\centering
\includegraphics[width=0.95\columnwidth]{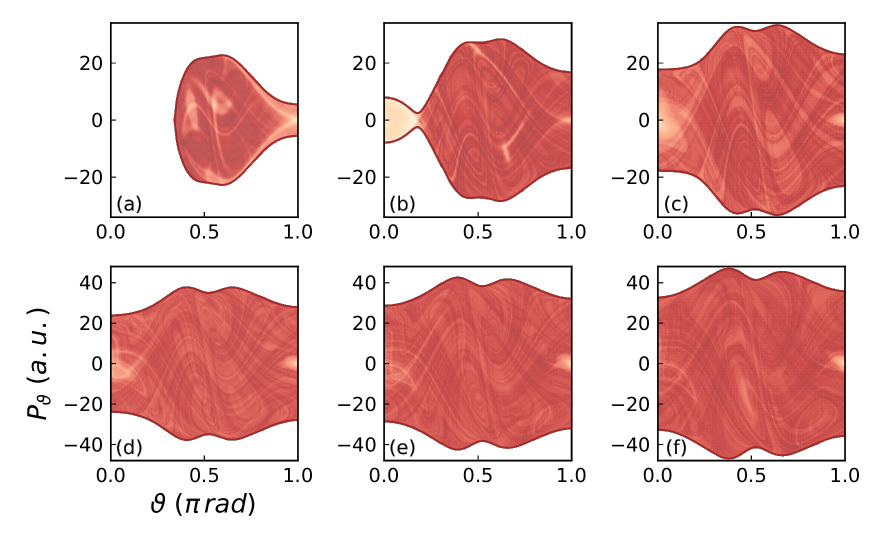}
\caption{
Same as Fig.~\ref{fig:LDs_smallE_t2d4} for
vibrational energies equal to
(a) $1000$\,cm$^{-1}$,
(b) $1500$\,cm$^{-1}$,
(c) $2000$\,cm$^{-1}$,
(d) $2500$\,cm$^{-1}$,
(e) $3000$\,cm$^{-1}$,
and
(f) $3500$\,cm$^{-1}$.
}
\label{fig:LDs_largeE}
\end{figure}

Figure~\ref{fig:LDs_largeE} shows the LD plots
for energies above the energetic barrier located
at~$\theta = \pi$\,rad,
which equals
\linebreak
$V(\pi, R_{\rm MEP}(\pi)) =939.6$\,cm$^{-1}$.
As shown in Fig.~\ref{fig:LDs_largeE}(a),
which corresponds to $E=1000$\,cm$^{-1}$,
most of the accessible phase space is located around the 
potential absolute minimum,
as in 
Figs.~\ref{fig:LDs_smallE_t2d4} and~\ref{fig:LDs_smallE_t1d5}.
However,
the motion is in this case
essentially governed by the invariant manifolds
associated with the stretch POs that are placed at the top
of the energetic barrier located at~$\theta=\pi$\,rad,
as discussed in detail elsewhere~\cite{Revuelta23}.
These manifolds,
and not those of the two POs introduced in the previous section,
appear automatically 
as continuous lines in the LD plots,
showing their importance for the global vibrational dynamics
of the molecule.
Notice, also, the presence of the corresponding homoclinic tangle
that is
responsible for the highly nonlinear dynamics.
Likewise,
recall that the energetically accessible area of the phase space 
is much smaller
in the neighborhood of the previous POs 
than around the potential minimum.
This difference lies on the fact that in the vicinity of the  barrier top
most of
the  mechanical energy is potential,
while around the potential minimum
it is mostly kinetic energy.

When the vibrational energy of the molecule surpasses the energy
$E=1494$\,cm$^{-1}$
of the main barrier that is located at~$\theta \approx 0.55 \pi$\,rad,
the motion can also take place around the relative minimum
located at~$\theta=0$\,rad.
This fact can be inferred from inspection of
Figs.~\ref{fig:LDs_largeE}(b)-(f),
where the LD plots for vibrational energies equal to
(b) 1500cm$^{-1}$, 
(c) 2000cm$^{-1}$, 
(d) 2500cm$^{-1}$, 
(e) 3000cm$^{-1}$, 
and
(f) 3500cm$^{-1}$ 
are shown.
As mentioned above within the discussion of Fig.~\ref{fig:LDs_largeE}(a),
the invariant manifolds that emerge from the
PO with~$\vartheta \approx \pi$\,rad,
become also visible in all cases considered,
which confims their 
dramatic impact on the
molecular dynamics.
Nevertheless,
the invariant manifolds 
associated with an aditional strecth PO located at~$\theta \approx 0.55\pi\,$rad
make the dynamics much more involved
due to the occurrence of a heteroclinic tangle
formed with the already existing invariant manifolds.
The changes in the vibrational energy yield also changes in the
stability of the PO, and in the invariant manifolds.
In fact,
as discussed in more detail below,
a  new region of stability around~$\theta = \pi$\,rad emerges
at~$E_{\rm bifur, 1}^{\rm SP} = 1343.2$\,cm$^{-1}$,
which manifests as an increasing-size light area 
around
$(\vartheta, P_\vartheta) = (\pi\,\textnormal{rad.}, 0\,\textnormal{a.u.})$
in the LD plots introduced in 
Figs.~\ref{fig:LDs_largeE}(b)-(f).

\subsubsection{Bifurcation scheme for high energies,
and the stabilization around the secondary energetic barrier}
\label{sec:stabilization}
In order to study in more detail the stabilization observed
around the secondary energetic barrier 
located at~$\theta=\pi\,$rad,
we analyze here the stability of the bifurcating PO
that is found in that region,
and the phase-space evolution in its neighboughhood
when the vibrational energy is tuned.

\begin{figure}[t!]
\centering
\includegraphics[width=0.95\columnwidth]{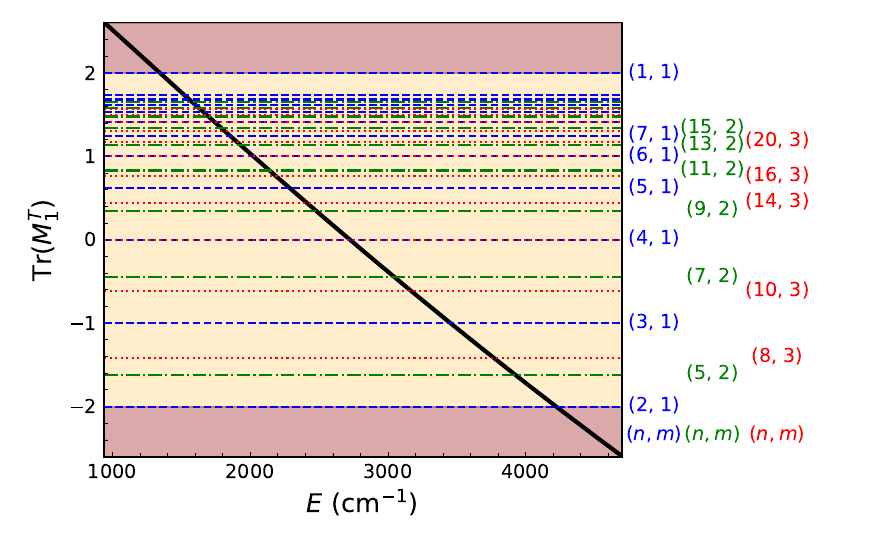}
\caption{Same as Fig.~\ref{fig:tr_centalPOs}
for the stretch periodic orbit that is
found at~$\theta \approx \pi$\,rad,
which is stable within the range~$1343.2$cm$^{-1}<E<4220.6$cm$^{-1}$.
}
\label{fig:tr_theta=pi}
\end{figure}

First,
in order to identify the precise value of the bifurcation energy,
where the stability region emerges,
we show in Fig.~\ref{fig:tr_theta=pi} the value of the trace
of the monodromy matrix of the PO 
located at~$\theta \approx \pi\,$rad
as a function of the vibrational energy.
Here, 
we have also marked with horizontal lines the positions
where the bifurcation condition~\eqref{eq:bifur} is fulfilled
along with some of the~$n$ and~$m$ indices.
As can be seen, 
the value of~$\Tr\left (\mathcal{M}_1^\trans \right)$
monotonically decreases with the energy
within the considered range.
At the saddle-point energy (939.65cm$^{-1}$)
the trace of the monodromy 
matrix is~$\Tr\left (\mathcal{M}_1^\trans \right) \approx 2.61$,
a value that is larger than~2,
so, 
acoording to the theory outlined in Sec.~\ref{sec:bifur},
it is associated with an unstable PO.
For~$E_{\rm bifur, 1}^{\rm SP}=1343.2$cm$^{-1}$,
the unstable PO suffers a bifurcation and turns marginally stable as
$\Tr\left (\mathcal{M}_1^\trans \right) = 2$.
At this point,
both eigenvalues~\eqref{eq:mu}
of the monodromy matrix are equal to~1,
and the stable manifold degenerates
with the unstable manifold. 
For energies only slightly larger than~$E_{\rm bifur, 1}^{\rm SP}$,
three stretch POs are visible.
The central PO is stable,
while the two POs on its left- and right-hand sides
are unstable,
being organized in a~V-shape that is
symmetrical respect to the~$\theta=\pi$\,rad axis.
(Recall that only one of them is included in the fundamental domain
$0 \le \theta \le \pi$\,rad.)
All these POs correspond to resonances with~$(n, m)= (1, 1)$, 
in agreement with the results predicted by the bifurcation theory
shown in 
Table~\ref{table:bifur}
and 
Fig.~\ref{fig:tr_theta=pi}.

\begin{figure}[t!]
\centering
\includegraphics[width=0.95\columnwidth]{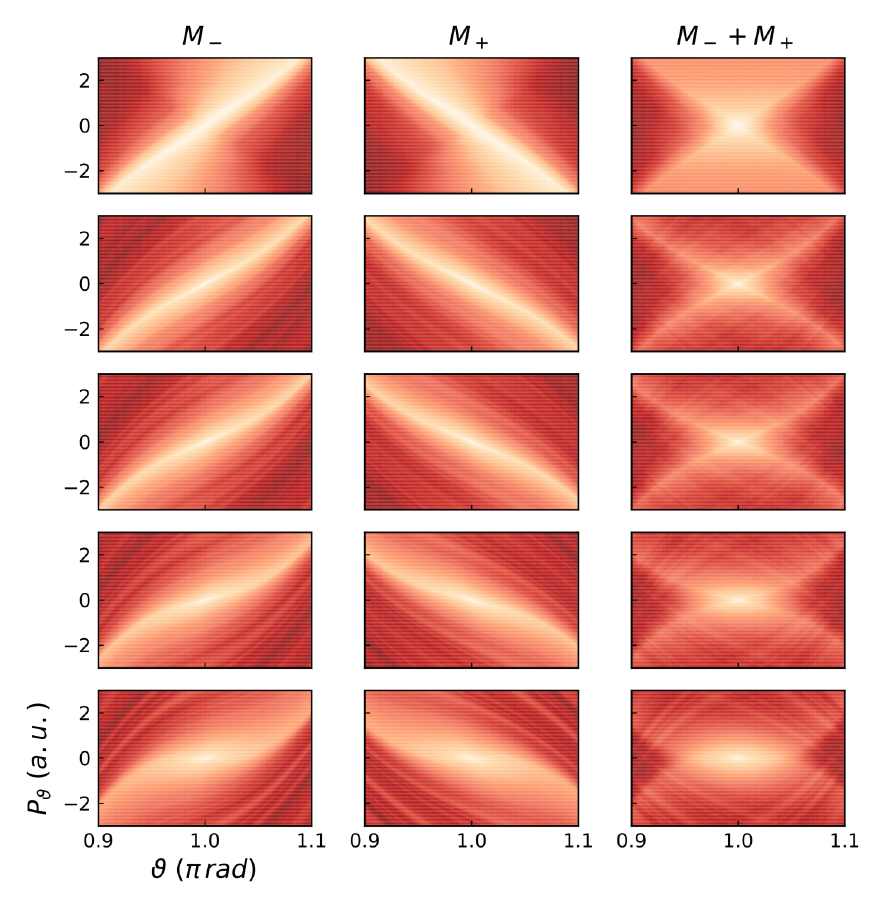}
\caption{Lagrangian descriptors~\eqref{eq:LD}
computed backwards  (left), ~$M_-$,
and forwards (middle),~$M_+$, in time,
and their sum,~$M_- + M_+$,
for~$\tau = 2 \times 10^4$\,a.u.,
and~$p=0.4$, and
vibrational energies equal to
(from top to bottom)
$940$\,cm$^{-1}$,
$1100$\,cm$^{-1}$,
$1200$\,cm$^{-1}$,
$E_{\rm bifur, 1}^{\rm SP}=1343.2$\,cm$^{-1}$,
and
$1500$\,cm$^{-1}$.
}
\label{fig:LDs_right_barrier_Ebifur1343p2}
\end{figure}

The imprint of the bifurcation on the phase-space structure
is also remarkably captured by the LDs,
as shown in
Fig.~\ref{fig:LDs_right_barrier_Ebifur1343p2},
where the LDs~\eqref{eq:LD} computed
solely backwards (left) and forward (middle) in time
along with their sum (right)
are shown.
Within the linear region,
the manifolds are alligned with the eigenvectors of the 
monodromy matrix.
The maximal slope of the eigenvectors, i.e.,
of the manifolds in the neighborhood of the PO,
is obtained at the saddle point energy, 
and then decreases.
Notice that for~$E=940$cm$^{-1}$
(a value only slightly larger than the potential energy
at the saddle point)
 the two manifolds
are almost linear lines that coincide with the diagonals of
the top panels in 
Fig.~\ref{fig:LDs_right_barrier_Ebifur1343p2}.
As the energy increases,
the central region of the manifolds gets distorted
and folds,
and the slope of the manifolds reduces,
as inferred from the comparison of the previous results
with those corresponding to~$E=1100$cm$^{-1}$
(second row),
and~$E=1200$cm$^{-1}$ (third row).
At the bifurcation energy,
the two manifolds coincide at the center with the
horizontal axis,
and degenerate,
as shown in the figure (fourth row).
At this point,
the original hyperbolic point in the LD plots
stabilizes and turns into a parabolic point.
As the energy further increases,
e.g. for 
$E=1500\,$cm$^{-1}$ shown in the bottom line of
Fig.~\ref{fig:LDs_right_barrier_Ebifur1343p2},
this parabolic point transforms into a central elliptic point
and two hyperbolic points at both sides of it.
A new stability region surrrounds the elliptic point.
The motion within this region is separated from the rest of the
phase space as it is confined by the invariant manifolds
that emerge from the two novel hyperbolic points.
These manifolds act as true separatrices for the dynamics.
From a global perspective,
the elliptic point 
corresponds to
a stable PO,
and two hyperbolic points to 
two unstable POs.

\begin{figure}[t!]
\centering
\includegraphics[width=0.95\columnwidth]{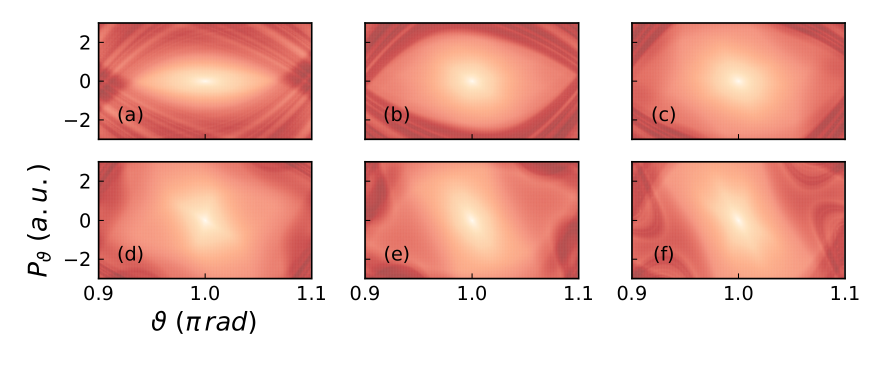}
\caption{
Same as Fig.~\ref{fig:LDs_smallE_t2d4} for
vibrational energies equal to
(a) $1600$\,cm$^{-1}$,
(b) $2000$\,cm$^{-1}$,
(c) $2400$\,cm$^{-1}$,
(d) $2800$\,cm$^{-1}$,
(e) $3200$\,cm$^{-1}$,
and
(f) $3600$\,cm$^{-1}$.
}
\label{fig:LDs_right_barrier_E1600-3600_t2d4}
\end{figure}

\begin{figure}[t!]
\centering
\includegraphics[width=0.95\columnwidth]{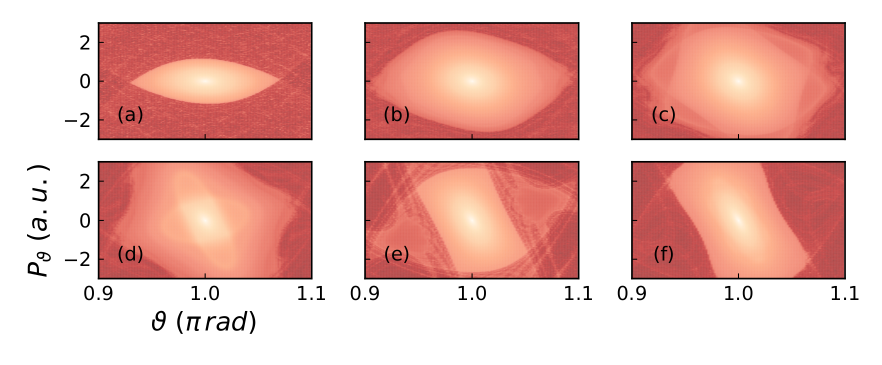}
\caption{Same as Fig.~\ref{fig:LDs_right_barrier_E1600-3600_t2d4}
for~$\tau = 10^5$\,a.u.
}
\label{fig:LDs_right_barrier_E1600-3600_t1d5}
\end{figure}

Figure~\ref{fig:LDs_right_barrier_E1600-3600_t2d4}
shows the sum of the forward and backwards LDs for
%
higher energies.
As can be seen,
the two hyperbolic points separate from the central elliptic point
with the energy, this increasing the size of the stability island
Similarly,
the corresponding unstable POs move appart from the
central stable PO.
In order to unveil the new structures that emerge 
within this region, we show in Fig.~\ref{fig:LDs_right_barrier_E1600-3600_t1d5}
the LD plots for the same energies but
computed up to a time of~$\tau = 10^5$\,a.u.
Here,
a more detailed view of the region surrounding the stable PO is
achieved.
Notice, in particular,
the 6 islands that are visible in 
Fig.~\ref{fig:LDs_right_barrier_E1600-3600_t1d5}(d)
due to the bifurcation~$(n, m) = (6, 1)=$ that takes place 
for a vibrational energy of~2022.6\,cm$^{-1}$,
and the 4 islands shown in 
Fig.~\ref{fig:LDs_right_barrier_E1600-3600_t1d5}(d)
that emerge when the~$(n, m) = (4, 1)$ bifurcation 
happens for a vibrational energy of~2723.7\,cm$^{-1}$
(see Table~\ref{table:bifur}).

\begin{figure}[t!]
\centering
\includegraphics[width=0.95\columnwidth]{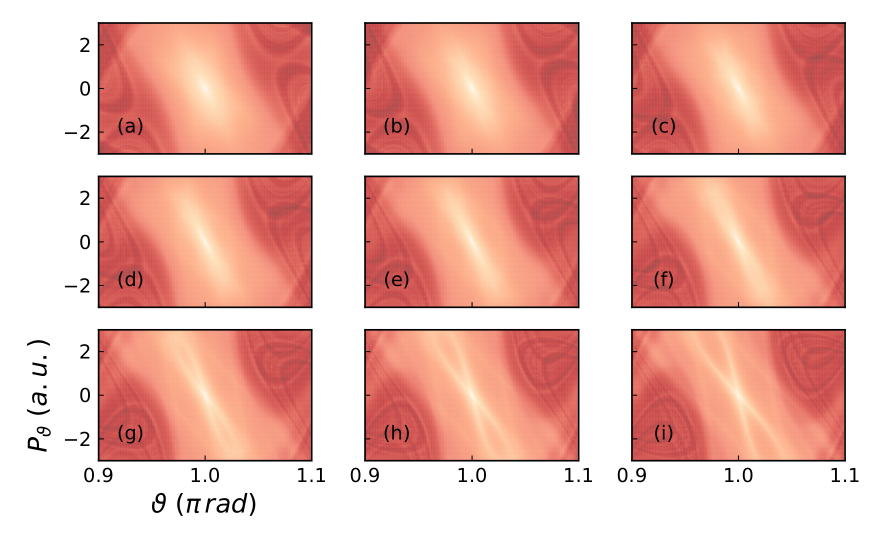}
\caption{Same as Fig.~\ref{fig:LDs_right_barrier_Ebifur1343p2}
for
vibrational energies equal to
(a) $3700$\,cm$^{-1}$,
(b) $3800$\,cm$^{-1}$,
(c) $3900$\,cm$^{-1}$,
(d) $4000$\,cm$^{-1}$,
(e) $4100$\,cm$^{-1}$,
(f) $4220.6$\,cm$^{-1}$,
(g) $4300$\,cm$^{-1}$,
(h) $4400$\,cm$^{-1}$,
and
(i) $4500$\,cm$^{-1}$.
}
\label{fig:LDs_right_barrier_Ebifur4220p6}
\end{figure}

For higher energies,
the PO remains stable until~$E_{\rm bifur, 2}^{\rm SP}=4220.6$cm$^{-1}$,
where it experiences another bifurcation
turning into a 1:2 unstable 
PO
because
$\Tr\left (\mathcal{M}_1^\trans \right) = -2$
(see Fig.~\ref{fig:tr_theta=pi}).
Notice the presence of a structure forming 
 an angle of~$\sim 80^\circ$
with the horizontal axis
for~$E<E_{\rm bifur, 2}^{\rm SP}$
in Figs.~\ref{fig:LDs_right_barrier_Ebifur4220p6}(a)-(e).
For~$E = E_{\rm bifur, 2}^{\rm SP}$,
these structures collapse, as shown 
in Fig.~\ref{fig:LDs_right_barrier_Ebifur4220p6}(f),
where a novel~$\infty$-shaped structure is observed.
This
$\infty$-shaped structure is formed by the invariant manifolds of the 
stretch PO that is located at the center of the plot.
Moreover,
it circunvents a stability region, where a 1:2 resonance 
is placed.
Likewise, as the energy further increases,
the size of this stable region increases, 
as shown in 
\ref{fig:LDs_right_barrier_Ebifur4220p6}(g),
\ref{fig:LDs_right_barrier_Ebifur4220p6}(h),
and
\ref{fig:LDs_right_barrier_Ebifur4220p6}(i)
related to
$E=4300$\,cm$^{-1}$,
$E=4400$\,cm$^{-1}$,
and
$E=4500$\,cm$^{-1}$,
respectively
(see Ref.~\cite{Parraga18} for further details). 

\begin{figure}[t!]
\centering
\includegraphics[width=0.95\columnwidth]{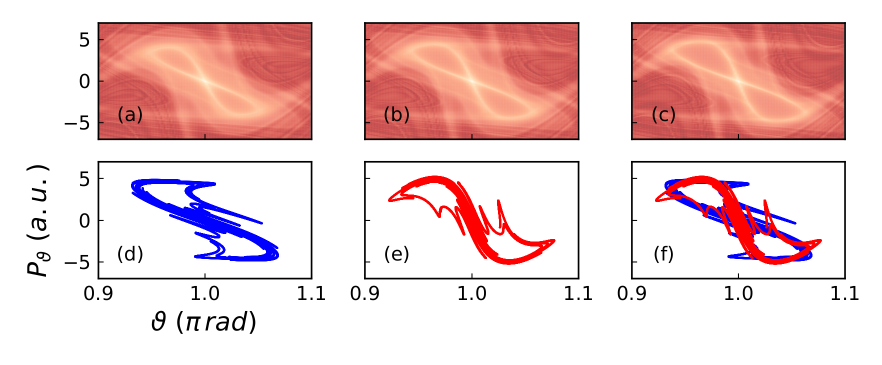}
\caption{Same as Fig.~\ref{fig:LDs_right_barrier_Ebifur1343p2}
for
vibrational energies equal to
(a) $4600$\,cm$^{-1}$,
(b) $4700$\,cm$^{-1}$,
and
(c) $4800$\,cm$^{-1}$,
%
and
(d) stable, and (e) unstable invariant
manifolds of the stretch periodic orbit
located at the top of the energetic barrier
found at~$\theta = \pi$\,rad
separated, and (f) superimpossed for
 $4800$\,cm$^{-1}$.
}
\label{fig:LDs_right_barrier_LD_manifolds}
\end{figure}

To conclude,
we show in Fig.~\ref{fig:LDs_right_barrier_LD_manifolds}
the LD plots for
vibrational energies equal to
(a) $4600$\,cm$^{-1}$,
(b) $4700$\,cm$^{-1}$,
and
(c) $4800$\,cm$^{-1}$.
Moreover,
in order to assess the accuracy of these computations,
we also present in the bottom panels of
Fig.~\ref{fig:LDs_right_barrier_LD_manifolds}
an explicit calculation of the invariant manifolds
by, first, diagonalizing of the corresponding
transversal monodromy matrix,
and, second, 
calculating a composite PSS~\eqref{eq:PSS}
related to the
time-propagation of a large set of
$10^4$ initial conditions located on the 
stable and unstable manifolds, respectively.
Figure
\ref{fig:LDs_right_barrier_LD_manifolds}(d)
shows the results for the stable manifold,
and
Fig.~\ref{fig:LDs_right_barrier_LD_manifolds}(e)
for the unstable manifold.
Notice the wild oscillations of the manifolds~\cite{Revuelta23},
caused by the strong anharmonicities in the PES,
wich are responsible for
the highly nonlinear molecular dynamics.
The agreement between the explicitly computed 
invariant manifolds
and the LDs results is better observed when comparing
Fig.~\ref{fig:LDs_right_barrier_LD_manifolds}(c)
with
Fig.~\ref{fig:LDs_right_barrier_LD_manifolds}(f),
where the previous manifolds have been superimpossed.

\section{Summary} \label{sec:summary}

In this work,
the phase-space structures of KCN molecular system have been analyzed
using Lagrangian descriptors.
For low energies,
the motion surrounds the minimum  of the potential energy surface.
In this situation,
the motion is regular due to the presence of invariant tori,
a fact that is detected by the Lagrangian descriptors
computed over a characteristic Poincaré surface of section
of the system.
As the energy increases,
the structure of the phase space changes dramatically.
On the one hand,
invariant tori breakdown as predicted by
the Kolmogorov-Arnold-Moser theorem. 
On the other hand,
a series of bifurcations take place.
We have observed that the Lagrangian descriptors
correctly account for all these changes if a sufficiently long
integration time is considered.

For energies 
above that of the  saddle point located 
at~$\theta = \pi$\,rad but below
the energy of the saddle point found at~$\theta=0.55\pi\,$rad,
the motion is essentially by the invariant manifolds that emerge
from the unstable 1:1 periodic orbits located at~$\theta \approx \pi$\,rad.
We arrive at this conclusion because
the plots of the Lagrangian descriptors
show up automatically as continuous lines 
these manifolds (and not others).
Furthermore,
these invariant manifolds also give rise to the emergence of a stable region
at the top of the energetic barrier.
At the center of this stable region,
a 1:1 resonance can be found,
which experience a series of bifurcations,
which have a clear imprint on the phase space.
We have shown that the stability analysis of the
central periodic orbit
correctly predicts the bifurcation energies.

For energies that surpass
the energy of the saddle point found at~$\theta=0.55\pi\,$rad,
the dynamics much more involved
due to the occurrence of a heteroclinic tangle
because of the interactions between 
the invariant manifolds 
associated with the periodic orbits that are located
at the top of the energetic barriers, 
surrounding the corresponding saddle points.

\section*{Acknowledgements} \label{sec:acknowledgements}
This work has been partially supported by the Grant PID2021-122711NB-C21
funded by
\linebreak
MCIN/AEI/10.13039/501100011033.
The authors also acknowledge computing
resources at the Magerit Supercomputer of the Universidad Politécnica de Madrid.


%

\bibliography{kcn_dnc2024}

\end{document}